\documentclass[11pt]{article}

\usepackage{amssymb, amsthm}
\usepackage{graphicx}
\usepackage{latexsym}
\usepackage{amsmath}
\usepackage{mathbbol}
\usepackage[utf8]{inputenc}
\usepackage{hyperref}

\begin{document}
\begin{titlepage}

\date{\today}
\begin{center}
{\bf\Large Semi-classical Schrödinger numerics in the residual representation}

\vskip1.5cm

{\normalsize Christoph N\"olle}\footnote{cnoelle.physics@gmail.com}


\end{center}

\vskip1.5cm

\begin{abstract} 
		The numerical treatment of quantum mechanics in the semi-classical regime is known to be computationally demanding, due to the highly oscillatory behaviour of the wave function and its large spatial extension. A recently proposed representation of quantum mechanics as a residual theory on top of classical Hamiltonian mechanics transforms a semi-classical wave function into a slowly-fluctuating, spatially confined \textit{residual} wave function. This representation is therefore well-suited for the numerical solution of semi-classical quantum problems. In this note I outline the formulation of the theory and demonstrate its applicability to a set of semi-classical scenarios, including a discussion of limitations. I work out the connection to established numerical approaches, such as the Gaussian beam approximation and the Gaussian wave packet transform by Russo and Smereka. A prototypical implementation of the method has been published as open-source software.		\end{abstract}
\end{titlepage}
\setcounter{tocdepth}{1}

 \tableofcontents

\section{Introduction}\label{sec:intro}

	Semi-classical quantum systems are known to be computationally demanding, due to strong oscillations of the wave function and the large domain in position space swept by the wave function. These imply that a grid-based discretisation requires a large grid size on the hand, and a very fine grid spacing on the other hand, making these problems very resource intensive. Several approaches are known to overcome these problems, including the WKB approximation, Gaussian beam approximations, and other methods related to the latter, described for instance in the two reviews \cite{JinMarkowichSparber11}, \cite{LasserLubich20} on the topic. A common property of these approaches is the need to impose a set of classical equations of motion on certain parameters of the problem, such as the Hamilton-Jacobi equation for the WKB approximation or the Hamiltonian equations of motion for the Gaussian beams. Whereas the majority of methods relies on explicit approximations, the so-called \textit{Gaussian wave packet transform} introduced by Russo and Smereka \cite{RussoSmereka13}, \cite{JinLiuRussoZhou20} starts from an exact reformulation of the Schrödinger equation. In this case only the unavoidable numerical errors lead to deviations from the exact solution.

	\bigskip

	In \cite{Noelle20} I have shown that the quantum model based on a wave function $\psi$ can be decomposed into a classical trajectory $c$ and a new wave function $\Phi$, such that all expectation values of observables $f$ also decompose into a sum of the classical expectation value on the trajectory $c$ and a quantum expectation value evaluated on $\Phi$. The choice of trajectory $c$ is arbitrary, in principle. However, for generic choices of $c$ the quantum expectation values will develop classical contributions under the time evolution, i.e., they may contain terms of order $\mathcal O(\hbar^0) = 1$. Only if $c$ satisfies Hamilton's equations and the initial value $\Phi_0$ is essentially confined to a cell in phase space of edge length $\sqrt\hbar$, the quantum expectation values will remain free of classical terms (they are then of order $\mathcal O(\hbar^{1/2})$). Furthermore, the confinement property of the wave function is then also preserved under the time-evolution, or more precisely, the scaling behaviour in the $\hbar \rightarrow 0$-limit is. This property is reminiscent of minimum uncertainty states, which satisfy the condition $\sigma_x\cdot \sigma_p = \hbar/2$ for the variances $\sigma_x$ and $\sigma_p$ in position and momentum space, respectively. 
	
	\bigskip

	The \textit{residual wave function} $\Phi$ being confined to a small cell in phase space, quite unlike the corresponding semi-classical wave function $\psi$ in the position representation, makes this representation also appealing to numerical applications. It implies that we only need a small grid size (because $\Phi$ is confined in the position dimensions) and a moderate grid resolution (because $\Phi$ is confined in momentum dimensions, and hence does not fluctuate strongly in position space). Hence, the two main problems for a numerical solution of the semi-classical Schrödinger equation seem to be resolved in the residual representation. 
	
	\bigskip
	
	There is a caveat to this argument, in that the confinement property of the residual wave function is really only a restriction on its scaling behaviour in the limit $\hbar \rightarrow 0$. Indeed, the example of a wave packet in the Morse potential at half the dissociation energy (Section \ref{sec:morse}) shows that the wave function may expand to such a large phase space volume that the advantages of the residual representation start to break down at some point. The effect is similar to the spreading of a free Gaussian wave packet. Nevertheless, as long as the residual wave function remains confined to a cell in phase space of edge length in the order of $\sqrt{\hbar}$ the method is applicable.

	\bigskip

	The \hyperref[sec:residual]{first section} of the paper recapitulates the residual representation, and the \hyperref[sec:Gaussian]{second one} discusses the relationship to the Gaussian beam approximation and the Gaussian wave packet transform. The latter in fact contains the residual representation and leads to an equivalent reformulation of the Schrödinger equation. The difference lies in a Gaussian factor in the wave function, which, as shown in this paper, is not actually required for the favourite numerical properties, although it is certainly convenient in many cases.

	\bigskip

	In the following sections the paper showcases numerical solutions in the residual representation for five sample scenarios, all one-dimensional. The \hyperref[sec:coherent]{first one} consists of the well-known coherent state solutions for the harmonic oscillator, which can be compared to their analytical solutions. The \hyperref[sec:quartic]{second example} consists of a quartic oscillator wave packet, and the \hyperref[sec:morse]{next one} of a wave packet in the Morse potential, at half the dissociation energy. As mentioned above, this example shows a limitation of the approach, when the wave packet starts to disperse in the flat region of the potential. Since the dispersed wave function is no longer confined to a small cell in phase space the advantages of the residual representation start to break down at this point, and the characteristic problems of semi-classical quantum dynamics show up again, i.e. strong oscillations and a large spatial extension of the wave function. 

	\bigskip

	The \hyperref[sec:tunneling]{next example} consists of a semi-classical wave packet that hits a potential barrier, where it is partly reflected and partly transmitted. The wave packet thus splits into two, moving into opposite directions. The classical trajectory, on the other hand, cannot split, it is either fully transmitted or fully reflected. The example illustrates that the residual formalism is not well suited for the propogation of the part of the wave function that lives on the other side of the potential barrier than the trajectory. A proposed workaround is to resort to the ordinary Schrödinger equation for the time duration of the splitting process, and to initialize two new trajectories and correspoding residual wave functions afterwards, tracing the time evolution of the two wave packets independently.

	\bigskip

	This example also demonstrates that the time evolution of the residual Schrödinger equation with the selected numerical integration scheme can violate unitarity, despite being based on the Crank-Nicolson scheme \cite{CrankNicolson47}, which is unitarity-preserving for the original Schrödinger equation.

	\bigskip 

	The \hyperref[sec:scattering]{final example} deals with a scattering problem for incoming waves of constant momentum, instead of localized wave packets as in the previous examples. It is shown that the residual representation can be beneficial in terms of boundary conditions, because it allows to transform a wave of constant, non-zero momentum into a constant function, allowing for the use of Neumann boundary conditions without reflection artifacts. 

	\bigskip
	
	The software prototype implementing the numerical integration scheme is described in an \hyperref[sec:appA]{appendix}. All numerical results displayed in the figures have been obtained by means of this application.

	\bigskip

	In this work the Planck constant $\hbar$, as well as position, time, mass and momentum, are considered dimensionless. The semi-classical regime in this convention is then defined by the condition $\hbar \ll 1$.

\section{The residual representation}\label{sec:residual}

	In this article we are dealing with non-relativistic point particles in a pure state, whose classical time evolution can be modeled in terms of a trajectory in phase space solving Hamilton's equations, and whose quantum evolution can be modeled by a wave function $\psi(t)$ solving the (time-dependent) Schrödinger equation: 
	
	\begin{align}\label{introSchroedinger}
		i\hbar \partial_t \psi (t) &= \hat H \cdot \psi(t) \\
			&= \sum_{k=0}^\infty \frac 1{k!} \big(\partial_{\alpha_1}  \dots \partial_{\alpha_k} H\big)(0) \hat y^{\alpha_1}\dots \hat y^{\alpha_k} \cdot \psi(t). \nonumber
	\end{align}
	
	Here, $H$ is the Hamiltonian function of the system, understood as a function of $2n$ phase space variables $y^1, \dots, y^{2n}$ and the term $\big(\partial_{\alpha_1}  \dots \partial_{\alpha_k} H\big)(0)$ denotes the evaluation of a k-th partial derivative of $H$ in the origin $0$ of phase space. Implicit summation over the indices $\alpha_1,\dots,\alpha_k$ from 1 to $2n$ is understood, and the $\hat y^{\alpha}$ are the canonical operators, usually written as $\hat q^i$ and $\hat p_j$ for $i,j=1,\dots, n$. They satisfy the canonical commutation relations
	\begin{equation}\label{introCanonicalCommutationRelation}
		[\hat y^\alpha,\hat y^\beta]=i\hbar\omega^{\alpha\beta},
	\end{equation}
	where $\omega$ is the symplectic form, concretely $(\omega^{\alpha\beta})_{\alpha,\beta = 1,\dots,2n} = \left(\begin{array}{cc}
					0_{n\times n} & \mathbb 1_{n\times n} \\
					-\mathbb 1_{n\times n} & 0_{n\times n}
			  \end{array} \right)$. Since the partial derivatives of the Hamiltonian in \eqref{introSchroedinger} are symmetric, we are using the \textit{Weyl} (or \textit{symmetric}) \textit{ordering convention} for the Schrödinger equation.

	\bigskip

	Now let $c$ be an arbitrary trajectory in phase space, and define the time-dependent unitary operator $U(t)$:

	\begin{equation}\label{introUnitaryDef}
		U(t) := U_{c(t)} = \exp \Big[ \frac i\hbar \int_0^t \Big\{H(c(\tau)) + \frac 12 \omega_{\alpha\beta} (\partial_\tau c^\alpha)(\tau) c^\beta(\tau) \Big\}d\tau \Big] \cdot \exp \Big[ -\frac i\hbar \omega_{\alpha\beta}c^\alpha (t)\hat y^\beta \Big].
	\end{equation}

	Here, the first factor is just a time-dependent phase (it removes the zeroth-order term in the $\hat y^\alpha$ in the residual Schrödinger equation introduced below). The second factor is a so-called \textit{Weyl operator}. The time-dependence in the Weyl operator via $c(t)$ is motivated by the role of the Weyl operator as a parallel-transport operator for Fedosov's connection used in deformation quantization \cite{Fedosov94}, \cite{Noelle20}. 

	\bigskip

	Define the \textit{residual wave function} $ \Phi (t) := U(t) \psi(t)$ and \textit{residual operator} $ \tilde f (t) := U(t) \hat f U(t)^{-1}$ for any observable $f$. It turns out that $\tilde f$ can be obtained from the Tayler expansion of $f$ in the phase space point $c(t)$:

	\begin{equation}
		\tilde f(t) = \sum_{k=0}^\infty \frac 1{k!} \big(\partial_{\alpha_1}  \dots \partial_{\alpha_k} f\big)\big(c(t)\big) \hat y^{\alpha_1}\dots \hat y^{\alpha_k}. 
	\end{equation}

	The resulting \textit{residual Schrödinger equation} satisfied by $\Phi$ reads:

	\begin{equation}\label{introSchroedingerResidual}
		i\hbar \partial_t  \Phi (t) = \bigg\{\Big[\partial_\alpha H(c(t)) - \omega_{\alpha\beta} \partial_t c^\beta(t)\Big] \hat y^\alpha  +\sum_{k=2}^\infty \frac 1{k!} \big(\partial_{\alpha_1}  \dots \partial_{\alpha_k} H\big)(c(t)) \hat y^{\alpha_1}\dots \hat y^{\alpha_k}  \bigg\} \Phi(t).
	\end{equation}

	For constant $c = 0$ we recover the ordinary Schrödinger equation (\ref{introSchroedinger}), up to a zeroth order term. If, on the other hand, $c$ satisfies the classical Hamiltonian equations
	
	\begin{equation}\label{CondHamilton1}
		\partial_t c^\alpha(t) = \omega^{\alpha\beta} \partial_\beta H(c(t)),
	\end{equation}
	
	then \eqref{introSchroedingerResidual} simplifies to

	\begin{equation}\label{introSchroedingerResidualOnshell}
		i\hbar \partial_t \Phi (t) = \bigg[ \sum_{k=2}^\infty \frac 1{k!} \big(\partial_{\alpha_1}  \dots \partial_{\alpha_k} H\big)(c(t)) \hat y^{\alpha_1}\dots \hat y^{\alpha_k}  \bigg] \Phi(t).
	\end{equation}

	The vanishing of the first order term in the canonical operators $\hat y^\alpha$ in (\ref{introSchroedingerResidualOnshell}) as opposed to the more general residual equation (\ref{introSchroedingerResidual}) or the original Schrödinger equation (\ref{introSchroedinger}) has some important consequences. It implies that the mapping from monomials $\hat y^{\alpha_1}\dots \hat y^{\alpha_k}$ to expectation values $\langle \Phi | \hat y^{\alpha_1}\dots \hat y^{\alpha_k} \Phi \rangle$, considered as functions of $\hbar$, remains filtration-preserving under the time-evolution, i.e. monomials of order $k$ have expectation values of order $\mathcal O(\hbar^{k/2})$ (if this is true for the initial wave function $\Phi(t_0)$) \cite{Noelle20}.
	 
	In particular, this implies that the quantum expectation values of an observable $f$:

	\begin{equation}\label{introExpectationValueClassicalExpansion1}
		\langle \hat f \rangle _{\psi(t)} = \langle \tilde f(t) \rangle _{\Phi(t)} = f(c(t)) + (\partial_\alpha f)(c(t)) \langle \hat y^\alpha \rangle_{\Phi(t)} + \dots,
	\end{equation}

	split into the classical part $f(c(t))$ plus quantum corrections at order $\mathcal O(\hbar^{1/2})$, so that in the classical limit $\hbar \rightarrow 0$ they are fully determined by the classical expectation values $f(c(t))$.
	Furthermore, it suggests that the residual Schrödinger equation (\ref{introSchroedingerResidualOnshell}) is well-suited for numerical applications, since the result could be understood such that $\Phi$ remains essentially confined to an interval of size $\hbar^{1/2}$ in all position dimensions, and likewise in the momentum dimensions, which means that it will not develop strong fluctuations in the position representation. Strictly speaking, however, the results only restrict the scaling behaviour of the expectation values with respect to variations of $\hbar$, whereas the behaviour at fixed $\hbar$, which is of interest to numerical applications, is not determined. We will encounter an example where the wave function indeed starts to spread beyond the small cell it is confined to initially in Section \ref{sec:morse} on the Morse potential, showing that this is indeed a practical concern. 
	
	\bigskip

	In the following, we will now focus on the position representation, i.e., we consider a position-dependent wave function $\psi=\psi(t,\vec x)$ for $\vec x\in \mathbb R^n$, and realize the canonical operators $\hat y^\alpha$ as follows:

	\begin{equation} \label{canonicalOperators}
		\hat q^j\psi(\vec x)= x^j\psi(\vec x), \qquad 
		\hat p_k\psi(\vec x) = \frac \hbar i\frac \partial{\partial x^k}\psi(\vec x).
	\end{equation}

	Furthermore, we assume from now on that $H$ has the form $H(\vec q,\vec p) = \frac{\vec p^2}{2m} + V(\vec q)$ for $\vec q, \vec p\in \mathbb R^n$, and in a slight abuse of notation we write $c(t) =(\vec q(t), \vec p(t))$, or simply $c=(\vec q, \vec p)$. The residual Schrödinger equation (\ref{introSchroedingerResidual}) becomes

	\begin{align}
		i\hbar \partial_t \Phi(t, \vec x) 
			&= \bigg\{ \frac{\hat p^2}{2m} + \Big[\frac {\vec p}m - \dot{\vec q}\Big] \cdot \hat{\vec p} + \Big[\vec\nabla V(\vec q) + \dot{\vec p}\Big]\cdot \hat{\vec q} \\
			& \qquad\qquad\qquad\qquad+ \sum_{k=2}^\infty \frac 1{k!} \big(\partial_{j_1}  \dots \partial_{j_k} V\big)(\vec q) \hat q^{j_1}\dots \hat q^{j_k}  \bigg\} \Phi(t, \vec x)  \nonumber \\
		&= \bigg\{ -\frac{\hbar^2}{2m}\Delta_x + \frac \hbar i\Big[\frac {\vec p}m - \dot{\vec q}\Big]\cdot \vec\nabla_x + \Big[\vec\nabla V(\vec q) + \dot{\vec p}\Big]\cdot \vec x \\
			& \qquad\qquad\qquad\qquad+ \sum_{k=2}^\infty \frac 1{k!} \big(\partial_{j_1}  \dots \partial_{j_k} V\big)(\vec q) x^{j_1}\dots x^{j_k}  \bigg\} \Phi(t, \vec x) \nonumber
	\end{align}

	In order to analyze the $\hbar$-dependence we substitute
	\begin{equation}
		\xi^j = \frac{x^j}{\sqrt\hbar}
	\end{equation}
	and bring all occurrences of $\hbar$ to the right-hand side:
	\begin{align}
		i\partial_t \Phi(t, \vec \xi) &=  \bigg\{ -\frac{1}{2m}\Delta_\xi + \frac i{\sqrt\hbar} \Big[\dot{\vec q} - \frac {\vec p}m \Big]\cdot \vec\nabla_\xi + \frac 1{\sqrt\hbar}\Big[\dot{\vec p} + \vec\nabla V(\vec q)\Big]\cdot \vec \xi \nonumber\\
		& \qquad \qquad+ \sum_{k=2}^\infty \frac {\hbar^{(k-2)/2}}{k!} \big(\partial_{j_1}  \dots \partial_{j_k} V\big)(\vec q) \xi^{j_1}\dots \xi^{j_k}  \bigg\} \Phi(t, \vec \xi) \label{SchroedingerPosGeneric}
	\end{align}
	In the case that $c$ satisfies Hamilton's equations, this simplifies to
	\begin{align}\label{SchroedingerPosHamilton}
		i\partial_t \Phi(t, \vec \xi) &=  \bigg\{ -\frac{1}{2m}\Delta_\xi + \sum_{k=2}^\infty \frac {\hbar^{(k-2)/2}}{k!} \big(\partial_{j_1}  \dots \partial_{j_k} V\big)(\vec q) \xi^{j_1}\dots \xi^{j_k}  \bigg\} \Phi(t, \vec \xi) 
	\end{align}

	Compare to the original Schrödinger equation (\ref{introSchroedinger}), which in terms of $\xi$ reads:

	\begin{align}\label{SchroedingerInXi}
		i\partial_t \psi(t, \vec \xi) &=  \bigg\{ -\frac{1}{2m}\Delta_\xi + 
		\frac {V(0)}\hbar+ \frac 1{\sqrt\hbar}\big(\vec\nabla V(0)\big)\cdot\vec \xi +
		\sum_{k=2}^\infty \frac {\hbar^{(k-2)/2}}{k!} \big(\partial_{j_1}  \dots \partial_{j_k} V\big)(0) \xi^{j_1}\dots \xi^{j_k}  \bigg\} \psi(t, \vec \xi) 
	\end{align}

	The first order term $\frac 1{\sqrt\hbar}\big(\vec\nabla V(0)\big)\cdot\vec \xi$ is critical here due to the $1/{\sqrt\hbar}$-dependence. In the semi-classical setting $\hbar \ll 1$ it can cause the highly-oscillatory behaviour of the wave function. Equation (\ref{SchroedingerPosHamilton}) on the other hand does not contain terms with inverse $\hbar$-dependencies, and is typically better behaved in the semi-classical regime. In principle, the first order term in \eqref{SchroedingerInXi} can also be eliminated by setting the origin of the $q$-coordinates to a minimum of the potential, so that $\vec\nabla V(0) = \vec 0$. However, if the wave function is not localized near this minimum initially, a macroscopic grid of the order $\hbar^{-1/2}$ will be required for the numerical solution.

	\bigskip

	While \eqref{SchroedingerPosHamilton} is useful for analyzing the $\hbar$-dependency, in its exact form it requires evaluating an infinite number of derivatives of the potential, unless the latter is polynomial. For numerical applications it is therefore usually more convenient to work with the original $x$-variable instead of $\xi = x / \sqrt{\hbar}$. Undoing the Tayler expansion of $V$, the equation becomes

	\begin{equation}\label{SchroedingerPosHamiltonClosed}
	i\partial_t \Phi(t, \vec x) =  \bigg\{ -\frac{\hbar}{2m}\Delta_x + \frac 1\hbar \Big[ V(\vec q + \vec x) - V(\vec q) - \vec x \cdot \vec\nabla V(\vec q) \Big] \bigg\} \Phi(t,\vec x).
	\end{equation}

	In this equation the term

	\begin{equation}\label{SchroedingerResEffectivePotential}
		V_\text{eff} (\vec x, t) = V\big(\vec q(t) + \vec x\big) - V\big(\vec q(t)\big) - \vec x \cdot \vec\nabla V\big(\vec q(t)\big)
	\end{equation}

	serves as a time-dependent effective potential. 

	\bigskip

	For reference, the transformation $U(t)$ from \eqref{introUnitaryDef} in our scenario acts as 

	\begin{equation}
		U(t)\cdot \psi(\vec x) = \exp \Big[ \frac i\hbar \int_0^t \Big\{ V(q(\tau)) - \tfrac 12 \vec q(\tau)\cdot \vec\nabla V(\vec q(\tau)) \Big\} d\tau \Big] e^{-\frac i\hbar \vec p\cdot\Big(\vec x + \tfrac 12 \vec q\Big)} \psi\big(\vec x + \vec q\big),
	\end{equation}

	if $c$ is Hamiltonian.

\section{Gaussian beams and Gaussian wave packet transform}\label{sec:Gaussian}

	Equation \eqref{SchroedingerPosHamilton} is an exact reformulation of the Schrödinger equation, not involving any approximations. It provides an expansion in orders of $\sqrt\hbar$:

	\begin{equation}\label{SchroedingerExpansionHbar}
		i\partial_t \Phi(t, \vec \xi) =  \bigg\{ -\frac{1}{2m}\Delta_\xi + \frac 12 \partial_j\partial_k V(\vec q) \xi^j \xi^k + \frac {\hbar^{\frac 12}}6 \partial_j\partial_k \partial_m V(\vec q) \xi^j \xi^k\xi^m + \mathcal O(\hbar^1) \bigg\} \Phi(t, \vec \xi).
	\end{equation}

	A zeroth-order approximation leads to a harmonic equation, with $\hbar$ factored out entirely:

	\begin{equation}\label{phiSchroedingerHarmonic}
		i\partial_t \Phi(t, \vec \xi) =  \bigg\{ -\frac{1}{2m}\Delta_\xi + \frac 12 \partial_j\partial_k V(\vec q) \xi^j \xi^k \bigg\} \Phi(t, \vec \xi).
	\end{equation}

	Note that the potential term is still time-dependent here via $\vec q = \vec q(t)$. A particular class of solutions to \eqref{phiSchroedingerHarmonic} is obtained from the \textit{Gaussian beam} ansatz:

	\begin{equation}\label{GaussianWavePacket}
		\Phi(t,\vec\xi) = A(t) \cdot \exp\bigg[\frac i2 \vec \xi^T \cdot M(t) \cdot \vec\xi\bigg]
	\end{equation}
	for complex $A(t)$ and some complex symmetric matric $M(t)$ with positive definite imaginary part. It can be readily verified that equation \eqref{phiSchroedingerHarmonic} becomes equivalent to

	\begin{equation}\label{GaussianWavePacketParamEq}
		\dot A = - \frac A{2m} \text{tr}(M), \qquad \dot M = -\frac{M^2}m - \nabla^2 V,
	\end{equation}
	with $\nabla^2 V$ denoting the Hessian of $V$, i.e. the matrix of second partial derivatives. In the original Schrödinger formulation, the ansatz \eqref{GaussianWavePacket} assumes the form

	\begin{equation}
		\psi(t, \vec x) = B(t) \cdot \exp\bigg\{  \frac i\hbar \Big[ \vec p\cdot(\vec x - \tfrac 12 \vec q)+ \frac 12 \big(\vec x -\vec q\big)^T\cdot M(t) \cdot\big(\vec x - \vec q\big)  \Big]  \bigg\}
	\end{equation}

	It is in this or similar forms that the Gaussian beam solutions are usually written (\cite{Heller76}, \cite{LasserLubich20}). Gaussian beams form the basis for many semi-classical numerical approaches, see for instance \cite{LasserLubich20}. The Gaussian wave packet \eqref{GaussianWavePacket} can be thought of as a generalization of the harmonic oscillator ground state to time-dependent and non-diagonal potentials. Analogues of the higher-order eigenstates have been constructed by Hagedorn, using generalized raising and lowering operators (\cite{Hagedorn98}, \cite{LasserLubich20}).

	\bigskip

	If we want to take into account higher orders in $\hbar$ in \eqref{SchroedingerExpansionHbar}, we are led to effective Hamiltonians with cubic and quartic time-dependent potentials. In this case the residual wave function $\Phi$ will depend explicitly on $\hbar$, not only via $\xi = x/\sqrt\hbar$.
	If we are looking for solutions to the general equation \eqref{SchroedingerPosHamilton} with initial conditions given by a Gaussian, then it may be useful to employ an ansatz for the full wave function consisting of a Gaussian beam as the lowest order solution times a correction factor for higher order terms, which we can expect to be slowly varying. This is the approach introduced by Russo and Smereka (\cite{RussoSmereka13}, \cite{JinLiuRussoZhou20}), called the \textit{Gaussian wave packet transform}:

	\begin{equation}\label{GaussianWavePacketTrafo}
		\Phi(t,\vec \xi) = A(t) \cdot \exp\bigg[\frac i2 \vec\xi^T \cdot M(t) \cdot \vec\xi\bigg] \cdot \kappa(t, \vec\xi)
	\end{equation}

	Assuming \eqref{GaussianWavePacketParamEq} to be satisfied and plugging the ansatz into \eqref{SchroedingerPosHamilton}, we obtain the following equation for $\kappa(t, \vec\xi)$:

	\begin{equation}\label{GaussianWavePacketTrafoEq}
		i\partial_t \kappa =- \frac 1{2m}\Big[\Delta \kappa + 2i (M\cdot \vec \xi)\cdot \vec\nabla \kappa\Big] + \sum_{k=3}^\infty \frac {\hbar^{(k-2)/2}}{k!} \big(\partial_{j_1}  \dots \partial_{j_k} V\big)(\vec q) \xi^{j_1}\dots \xi^{j_k} \kappa
	\end{equation}

	This equation is again fully equivalent to the original Schrödinger equation and the residual Schrödinger equation \eqref{SchroedingerPosHamilton}. Like \eqref{SchroedingerPosHamilton}, it provides a formulation better adapted to the semi-classical setting because the main oscillatory part of the wave function has been split off. It should be particularly useful when the initial wave function is of Gaussian form, so that the initial condition for $\kappa$ becomes $\kappa(t=0, \vec\xi) = 1$, whereas \eqref{SchroedingerPosHamilton} may be preferable for more general settings.

\section{Example 1: coherent states}\label{sec:coherent}

	In this first example we consider the harmonic oscillator in a single dimension of mass $m=1$ and frequency $\omega=1$:
	
	\begin{equation}
		H(q,p) = \frac12\big(p^2 + q^2\big).
	\end{equation}
	
	It is a very special case where we can actually construct semi-classical solutions analytically, the so-called coherent states. This allows us to compare numerical solutions to the exact solutions. Another special feature of quadratic Hamiltonians in general is the vanishing of the exponential term in the integral prefactor of our unitary transformation $U(t)$ (see \eqref{introUnitaryDef}):

	\begin{align}
		H(c) + \frac 12 \omega_{\alpha\beta} \dot c^\alpha c^\beta 
			&= H(c) - \frac 12 c^\alpha \partial_\alpha H(c) = 0, 
	\end{align}

	if $c$ satisfies Hamilton's equations \eqref{CondHamilton1}. The residual Schrödinger equation \eqref{SchroedingerPosHamilton} becomes:

	\begin{equation}\label{harmonicSchroedingerResdiual}
		i\partial_t \Phi(t,\xi) = \frac12\Big[-\Delta_\xi + \xi^2\Big]\Phi(t,\xi).
	\end{equation}

	It is another peculiar feature of the quadratic case that the dependence on the classical solution $c=(q(t),p(t))$ has completely disappeared from the right hand side, so irrespective of which classical solution we select, the residual Schrödinger equation remains the same. But of course, since $c$ enters into the transformation $U(t) = \exp(-\tfrac i\hbar \omega_{\alpha\beta}c^\alpha \hat y^\beta)$, the original wave function $\psi$ depends on the choice of trajectory. 

\begin{figure}
	\includegraphics[width=0.99\textwidth]{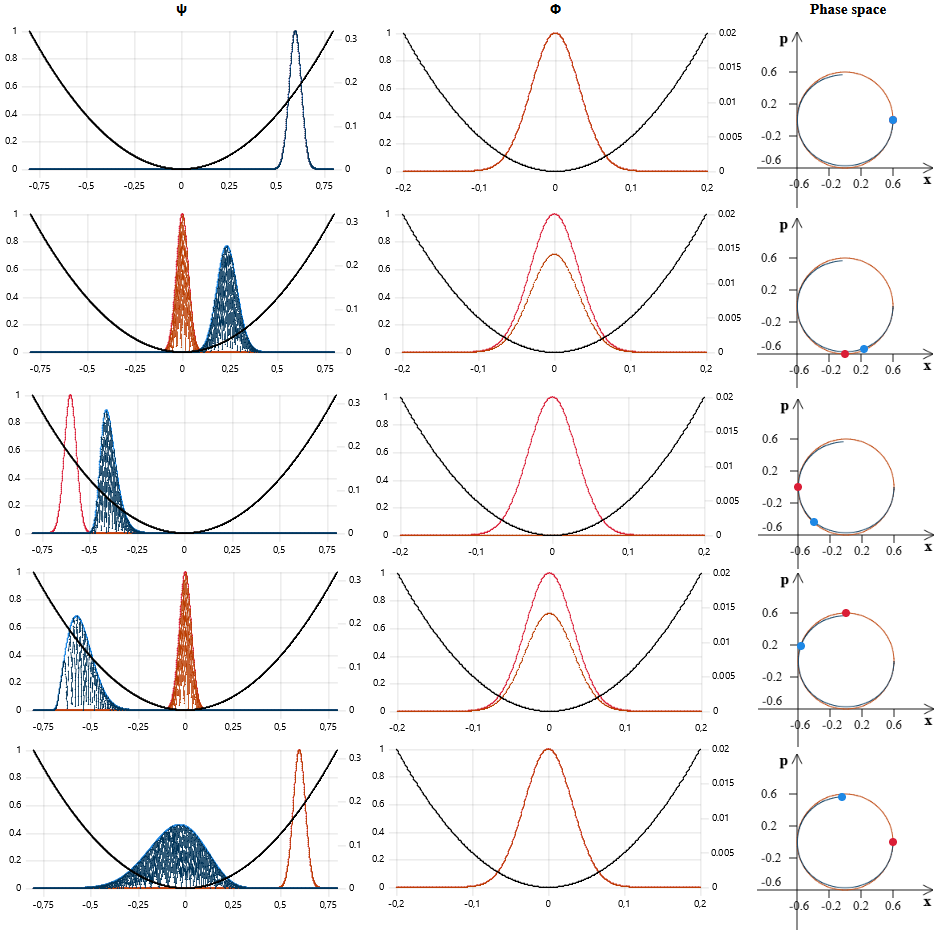}
	\caption{Numerically determined time evolution of a coherent state for a 1d harmonic oscillator, with every row displaying a snapshot at a fixed time step. The left column shows the wave function in the usual Schrödinger position representation, the middle column shows $\Phi$, the residual wave function. On the left, the blue wave function has been obtained by a direct integration of the Schrödinger equation, whereas the red one has been obtained from $\Phi$ shown in the middle by a Weyl transformation. For all wave functions, both the absolute value (solid line) and the absolute value of the real part (dashed line) are shown, they are indexed by the left y-axis. The right y-axis of the wave function plots applies to the potential $V(x)$, shown as a black parabola. The right column shows the phase space expectation values $(\langle \hat q\rangle, \ \langle\hat p\rangle)$ for the two wave functions.}
	\label{fig:ho-plots}
\end{figure}

	In the following we will focus on the ground state solution to \eqref{harmonicSchroedingerResdiual}, although the analysis equally applies to higher eigenfunctions of the stationary problem:

	\begin{equation}
		\Phi(t,\xi) = \exp\Big[-\frac 12 \Big( it +\xi^2 \Big)\Big]
	\end{equation}

	In the case $c=(0,0)$ we obtain $\psi(x) = \Phi(x)$, so $\psi$ actually models the ground state of the harmonic oscillator. If $c$ is a more general classical solution:

	\begin{equation}\label{OscTrajectory}
	c(t) = \exp\Big\{ \left(
					\begin{array}{cc}
					  0 & 1 \\
					  -1 & 0 \\
					\end{array}
				  \right)t
	 \Big\} c(0) = \left(\begin{array}{cc}
				 \cos\ t & \sin\ t \\
				 -\sin\ t & \cos\ t
			   \end{array} \right)c(0),
   \end{equation}

   then we obtain, writing $c(0) = (q_0,p_0)$:

   \begin{equation}
	 \psi(t,x) = e^{ i\alpha(t)}
	 \exp\Big[-\frac 1{2\hbar} \Big(x - \cos(t)q_0 -\sin(t)p_0 \Big)^2\Big].
   \end{equation}
   This is a coherent state solution, a Gaussian whose center follows the classical trajectory $c$. The phase is explicitly:
   \begin{equation}
  	e^{ i\alpha(t)} = e^{-\frac i2 (t + q_0p_0/\hbar)}
	\exp \Big[\frac i\hbar \Big(\sin^2(t) q_0p_o - \tfrac 12 \sin(t)\cos(t)[p_0^2-q_0^2]
		+ x\big[\cos(t)p_0 - \sin(t)q_0\big] \Big)\Big]
   \end{equation}

   Note that even though the wave function $\psi$ looks rather simple, for large values of $p_0, q_0$ (the semi-classical parameter range), the phase fluctuates strongly with $x$, implying that a direct numerical solution that needs to resolve the real and imaginary part of the wave function requires a fine grid-size. Simulating the evolution of $\Phi$ on the other hand is as simple as propagating the harmonic oscillator ground state.

\bigskip

Figure \ref{fig:ho-plots} shows a comparison of the numerically obtained wave functions, with one calculation done in the ordinary Schrödinger representation (blue wave function in the left column) and one in the residual Schrödinger representation based on the trajectory \eqref{OscTrajectory} (red wave function in the left column). The same grid settings have been used for both calculations. Parameters chosen were $\hbar=0.001$ and $(q_0, p_0) = (0.6, 0)$, which implies that the energy of our coherent state is $E = H(c(0)) + \frac {\hbar\omega}2 = 0.18 + 0.005 = 0.1805$, with the clasical contribution clearly dominating. Comparing the results with the analytical solution for the coherent state confirms that the calculation in the residual representation has a very small error, with the wave function being almost indistinguishable from the exact solution in the considered time interval. The calculation in the original Schrödinger representation, on the other hand, leads to a quickly accumulating error, with the wave function lagging behind the exact solution and spreading. In this case, the reason for the lagging is the insufficient grid resolution, which leads to slight truncations of the wave function oscillations and hence to a reduced momentum of the wave packet. This can also be observed in the phase space expectation values shown in the right column of Figure \ref{fig:ho-plots}, where the blue curve deviates slightly from the expected perfect circle, showing a reduced peak momentum.

\bigskip

The selected numerical time resolution is $\Delta t=  2 \pi/1000$ for the residual Schrödinger equation and $\Delta t_{\text{Ham}} = \Delta t / 100$ for Hamilton's equations. The snapshots shown in Figure \ref{fig:ho-plots} have been taken at timesteps 0, 250, 500, 750 and 1000, in units of $\Delta t$, showing exactly one classical oscillation period.

\section{Example 2: quartic oscillator}\label{sec:quartic}

	In this example we consider a Hamiltonian of the form

	\begin{equation}
		H(q,p) = \frac {p^2}2 + \frac {q^2}2 + \frac {q^4}6.
	\end{equation}

\begin{figure}
	\includegraphics[width=0.99\textwidth]{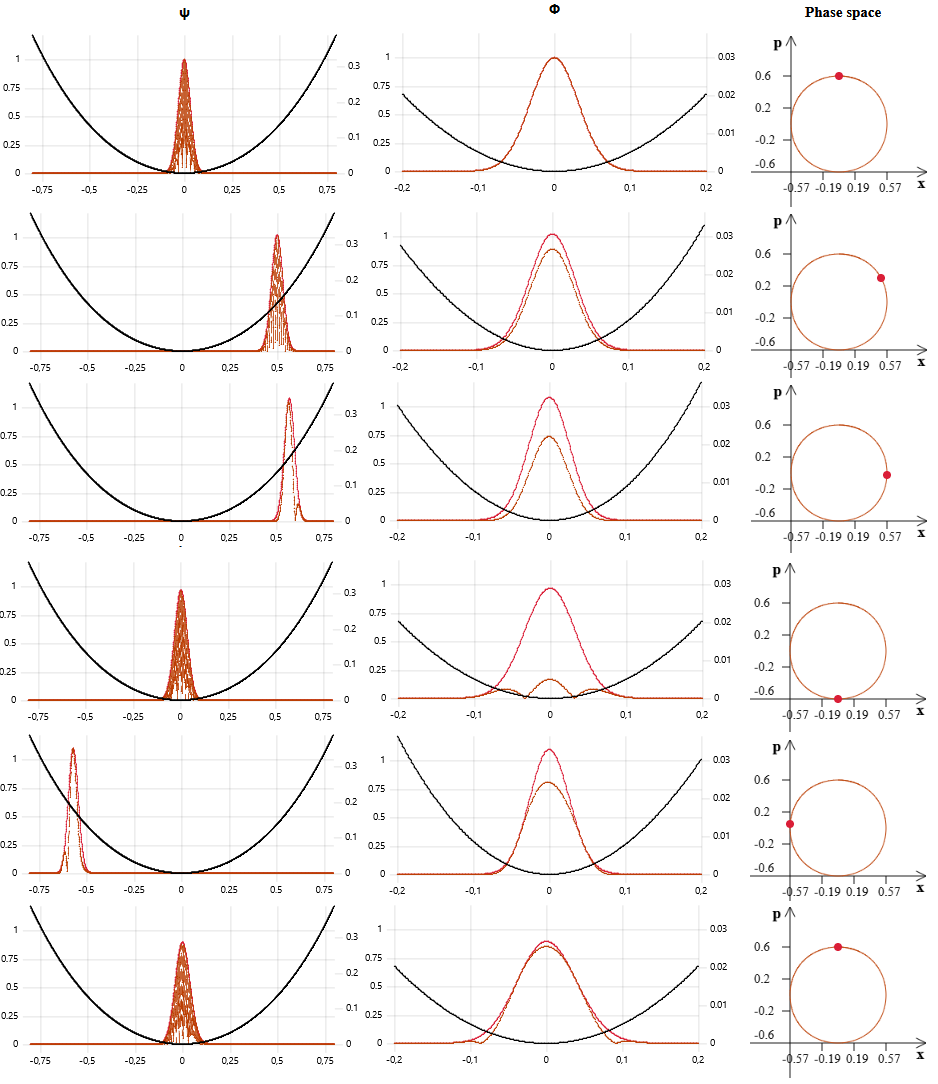}
	\caption{Time evolution of a wave packet for a 1d quartic oscillator. The left column shows the wave function in the usual Schrödinger position representation, the middle column shows $\Phi$, the residual wave function, and the right column shows the phase space expectation values.}
	\label{fig:quartic-plot}
	\end{figure}

	The initial base point in phase space is chosen as $(q_0,p_0) = (0,0.6)$ and the residual wave function is a Gaussian wave packet:

	\begin{equation}
		\Phi_0(x) = \exp\Big(- \frac{x^2}{2\hbar}\Big).
	\end{equation}

	The numerical solution obtained from the residual Schrödinger equation \eqref{SchroedingerPosHamiltonClosed} for $\hbar=0.001$ is shown in Figure \ref{fig:quartic-plot}. It is very similar to the coherent state solution of the harmonic oscillator (Figure \ref{fig:ho-plots}), except that the phase space trajectory becomes slightly non-circular. Note also that unlike the harmonic case the effective potential \eqref{SchroedingerResEffectivePotential} shown in the central column as the black solid line along with $\Phi(x)$ (in red) becomes truly time-dependent here. The selected time resolution is the same as for the harmonic oscillator example above, and snapshots shown in Figure \ref{fig:quartic-plot} are taken at 
	timestamps $0, 159, 240, 463, 705, 928$, in units of $\Delta t$ and from top to bottom.

\section{Example 3: Morse potential}\label{sec:morse}

	The 1-dimensional Morse potential is 

	\begin{equation}
		V(q) = D(1-e^{-aq})^2
	\end{equation}
	
	for parameters $a, D > 0$.
	It has a minimum in q = 0. The stationary Schrödinger equation for the Morse potential can be solved exactly and it has a finite number of bound states.
	The Morse potential can be used as a model for the binding energy of a diatomic molecule, where the coordinate $q$ corresponds to the relative distance between the atoms. The parameter $D$ is the dissociation energy in this picture.

	\bigskip

\begin{figure}
\centering
\includegraphics[width=0.6\textwidth]{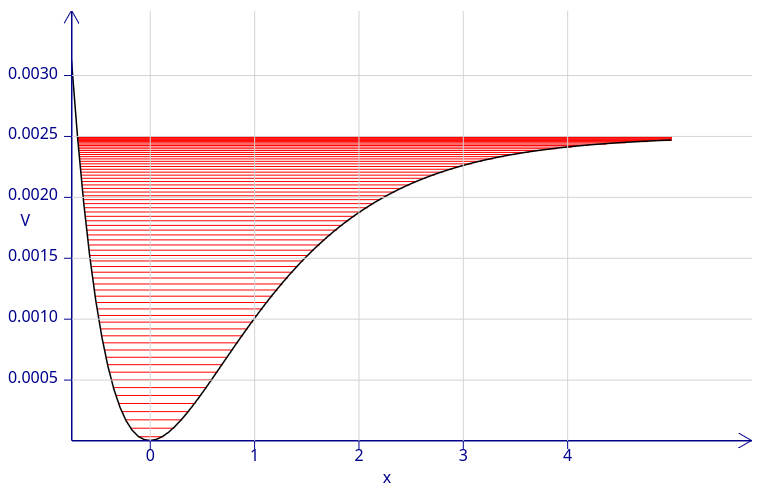}
\caption{Eigenvalues in the Morse potential.}
\label{fig:morse-eigenvalues}
\end{figure}

	Since $V''(0) = 2a^2D$, the harmonic approximation to the potential around 0 corresponds to a frequency
	\begin{equation}
		\omega^2 = \frac{2a^2D}{m}.
	\end{equation}
	The bound state energy levels of the Morse Hamiltonian are
	\begin{equation}
		E_n = \hbar \omega(n + \tfrac 12) - \frac {(\hbar \omega(n + \tfrac 12))^2}{4D}
	\end{equation}
	for integer $n\geq 0$, bounded by
	\begin{equation}
		n \leq \frac{\sqrt{2mD}}{a\hbar} - \frac 12.
	\end{equation}
	In the following we will consider an example for parameters 
	\begin{equation}
		a = 1, \qquad m=1, \qquad \hbar = 0.001, \qquad D = 5000  \frac{a^2  \hbar^2}{2 m}=0.0025,
	\end{equation}
	which implies that the maximum bound quantum number is $n_{max}=70$, and energies of the order of the dissociation energy $D$ can be considered to be more or less in the semi-classical regime. Figure \ref{fig:morse-eigenvalues} shows the eigenvalues for our setup.

	\bigskip

\begin{figure}
\includegraphics[width=0.99\textwidth]{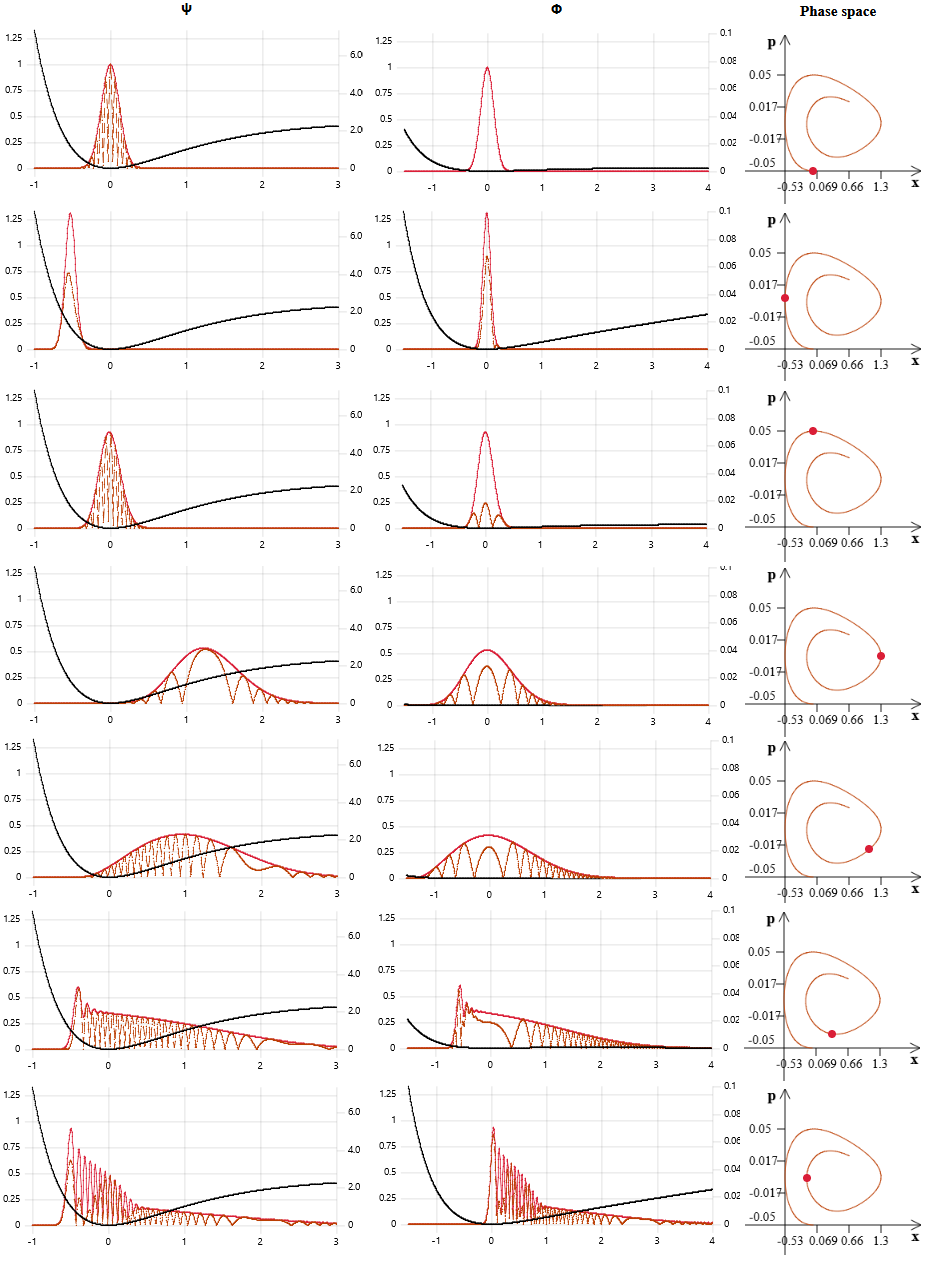}
\caption{Time evolution of a wave packet in a 1d Morse potential at half the dissociation energy.}
\label{fig:morse-plot}
\end{figure}

	We select the initial residual wave function as the ground state of the harmonic oscillator approximation at $q=0$:

	\begin{equation}
		\Phi_0(x) = \exp\Big(-\sqrt{\frac {mD}2}\frac a\hbar x^2\Big) 
	\end{equation}

	and start the classical trajectory at the phase space point $(q_0, p_0) = (0, -\sqrt{mD})$, whose contribution to the kinetic energy is exactly half the dissociation energy. Since the quantum contribution to the energy from $\Phi_0$ is small compared to the classical contribution, we are well in the confined region. Figure \ref{fig:morse-plot} shows the time evolution of the position wave function $\psi$ (left column), the residual wave function $\Phi$ (central column), and the phase space expectation values determined from $\Phi$ (right column). 

	\bigskip

	The first three to four frames show a time evolution similar to the harmonic and quartic oscillators, with the residual wave function $\Phi$ oscillating very little. But when the trajectory approaches its turning point at the flat side of the potential (right side in the plots), the effective potential \eqref{SchroedingerResEffectivePotential}, shown as the black solid line in the central column, becomes extremely flat (even assuming negative values, which are suppressed in the plot) and the wave function starts to disperse, simlarly to a free particle wave packet. The advantage of the residual representation starts to break down at this point. The inwards spiralling behaviour of the phase space expectation values seen in the lower plots is an artifact caused by the insufficient grid resolution for the strong oscillations of the wave function.

	\bigskip

	The quantum time step in these examples is chosen as $\Delta t = \frac{2\pi}{1000\, \omega}$ and the classical time step is $\Delta t_{\text{Ham}} = \Delta t/100$. Snapshots shown in Figure \ref{fig:morse-plot} are taken at 
	timestamps 0, 180, 350, 900, 1130, 1380, 1600, in units of $\Delta t$ and from top to bottom.

	\bigskip

\begin{figure}
\centering
\includegraphics[width=0.4\textwidth]{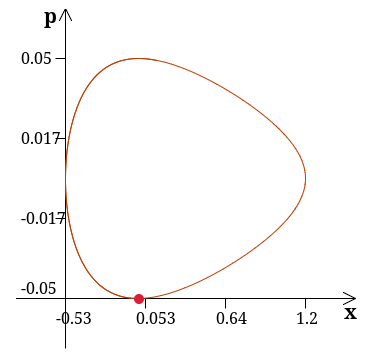}
\caption{Time evolution of a classical trajectory in a Morse potential at half the dissociation energy.}
\label{fig:morse-classical}
\end{figure}

	The corresponding classical trajectory is shown in Figure \ref{fig:morse-classical}. Its oscillation interval is roughly $1415\Delta t$.

\section{Example 4: reflection and transmission}\label{sec:tunneling}

	In the following example we consider the time-evolution of a semi-classical wave packet hitting a potential barrier, where for a limited time duration quantum effects become important to the dynamics. Concretely, they lead to a partial reflection of the wave function and hence to a splitting into a reflected and a transmitted wave packet. It should be rather obvious that the residual representation is not well suited for such a scenario, since the classical trajectory cannot split; it is either transmitted entirely or reflected entirely, depending on whether its energy exceeds the potential barrier or not. The residual scheme will work well for the part of the wave function that lives in the same region of position space as the classical trajectory, but will not be suitable for the part confined to the other region.

	\bigskip

\begin{figure}
		\includegraphics[width=0.99\textwidth]{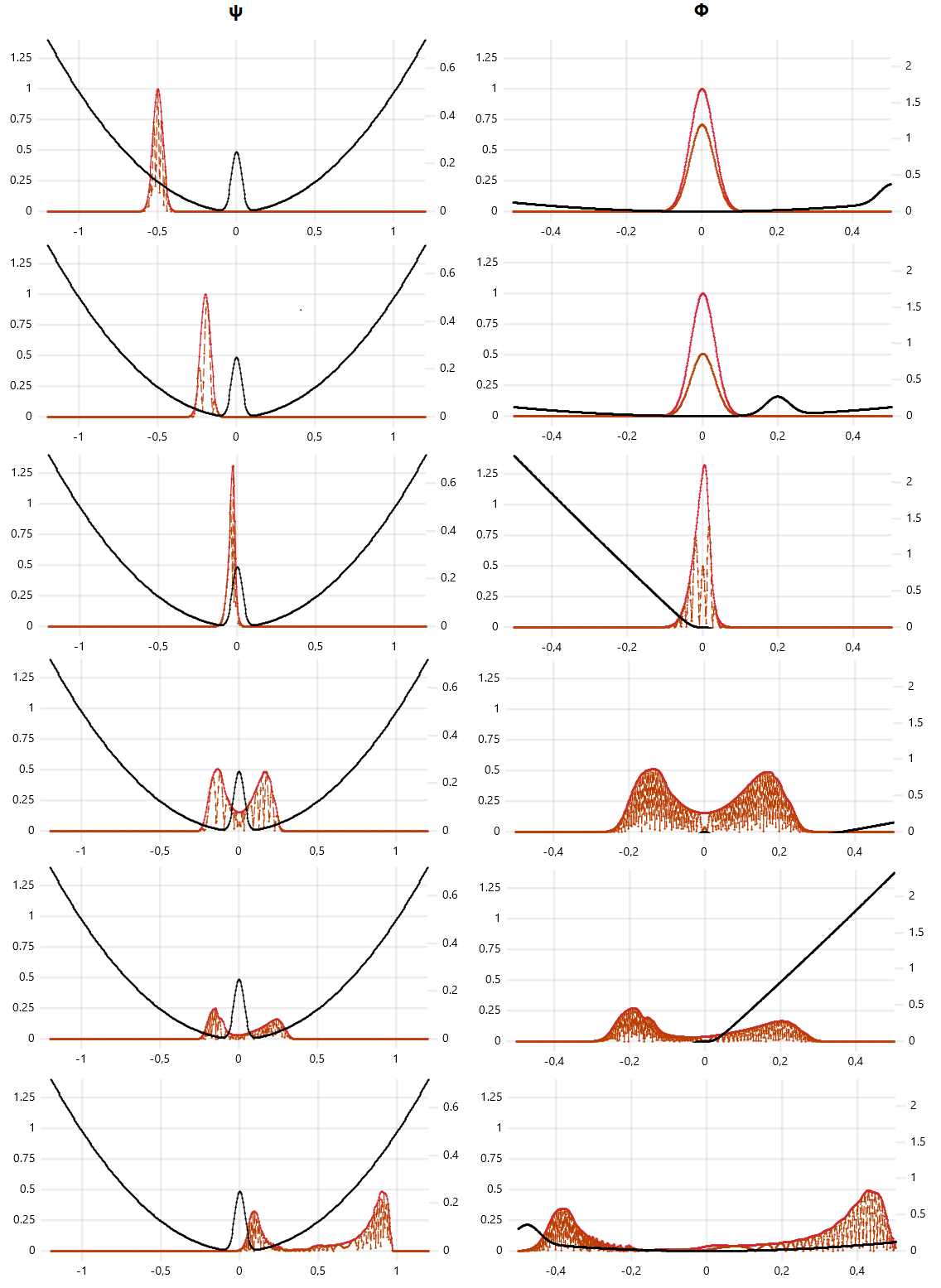}
		\caption{Numerical treatment of reflection and transmission of an incoming wave packet in the Schrödinger equation using the residual representation.}
		\label{fig:tunneling-plot-residual}
	\end{figure}

	For our example we choose a harmonic potential with an additional peak at the origin:

	\begin{equation}
		V(q) = \Big(1 + \frac{q^2}2\Big) \Big(1 + a \exp(-q^2/2b)\Big) - 1
	\end{equation}
	with $a,b > 0$. In the following we set $a=1/4$, $b=0.001$ and $\hbar=0.001$. The potential is displayed in the left column of Figure \ref{fig:tunneling-plot-residual} as the black solid line. The initial wave function $\psi(t_0)$ is based on a coherent state whose energy essentially equals the height of the barrier. This will lead to an effect that half of the wave packet is reflected back and the other half is transmitted past the barrier, when the wave function hits the latter. The classical trajectory in this examples traverses the barrier and continues on the other side.

	\bigskip

	Figure \ref{fig:tunneling-plot-residual} shows the result of a naive application of the residual integration scheme for our potential. The classical trajectory in this example starts on the left side of the barrier and traverses it to the right side between the 3rd and 4th frame shown. 	
	The wave function $\psi$ at first propagates as expected, splitting into two wave packets when it hits the barrier, with one packet being reflected (moving left) and the other one being transmitted (moving right). However, at a later timestamp, the reflected wave packet is \textit{dragged} to the other side of the barrier, as well (see last frame in Figure \ref{fig:tunneling-plot-residual}). This is obviously a numerical artifact, caused by the underlying classical trajectory being located beyond the barrier on the right hand side at this time. 
	
	\bigskip
	
	Another surprising effect is the strong non-unitarity of the time-evolution during the period of quantum dominance. In the transition from frame four to five in Figure \ref{fig:tunneling-plot-residual} the norm of the wave function decreases to roughly 30\% of its initial value, but then recovers upon the transition to the final frame to ca. 75\%. The numerical integration scheme used here is a straight-forward adapation to the residual setting of the Crank-Nicolson scheme for the original Schrödinger equation \cite{CrankNicolson47}. The latter is known to be unitarity-preserving, but apparently this is not the case for the residual Schrödinger equation.

	The time step is set to $\Delta t = \frac{2\pi}{1000}$ here (Hamiltonian time step $\Delta t_{\text{Ham}} = \Delta t/100$), and the frames shown are recorded after 250, 330, 370, 440, 470, 580 time steps, respectively, from top to bottom.

	\bigskip

\begin{figure}
\includegraphics[width=0.99\textwidth]{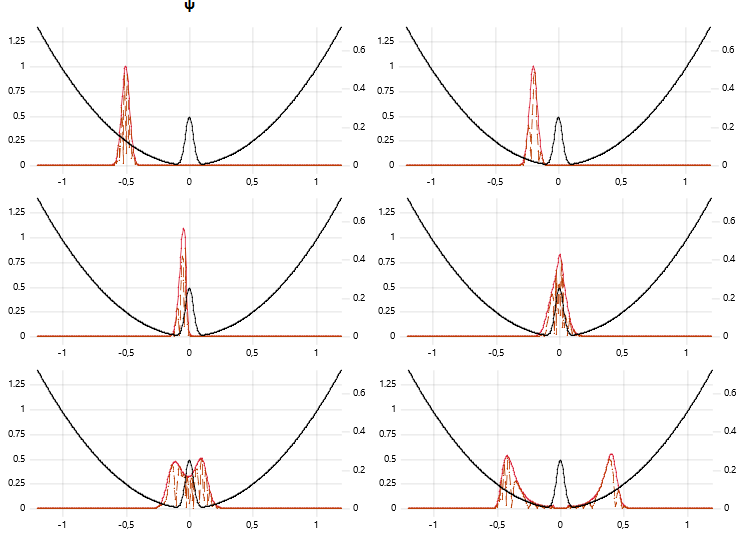}
\caption{Numerical treatment of reflection and transmission of an incoming wave packet in the Schrödinger equation using a combination of residual representations and the ordinary position representation. Timesteps increasing from top left to bottom right. The residual wave functions are not shown here, only the original wave function $\psi$.}
\label{fig:tunneling-plot-concat}
\end{figure}

	The problem can be overcome, for instance, by freezing the underlying phase space point for the duration of the quantum dominance, and splitting the problem into two afterwards, following two different classical trajectories corresponding to the two wave packets centers. Figure \ref{fig:tunneling-plot-concat} shows the result of such a splitting for our problem at hand. The timestamps of the frames displayed are the same as in Figure \ref{fig:tunneling-plot-residual}. The simulation starts with the residual representation, and after 358 integration steps (between frame 2 and 3) the phase space point is frozen at $(q,p)=(0,0)$, resorting to the classical Schrödinger equation for the following steps. After the wave function has split into two clearly separated wave packets $\psi_1$ and $\psi_2$, concretely after 558 integration steps or some time before the last frame, the problem is split into two residual representations, each following the classical trajectory determined by the expectation values of $\hat q$ and $\hat p$ in the respective Schrödinger wave functions $\psi_1$ and $\psi_2$. The displayed wave function $\psi$ in the last frame is the sum of the two Schrödinger wave functions obtained by individually back-transforming the residual wave functions.

	\bigskip

	As can be seen from the last frame of Figure \ref{fig:tunneling-plot-concat} the wave packets continue to live on opposite sites of the barriers. In fact, each wave packet individually propagates similarly to a coherent state in the harmonic potential. Unitarity of the time evolution is restored, as well.

\section{Example 5: a scattering problem}\label{sec:scattering}

	Finally, we consider an example of a scattering problem. In this case we select a classical trajectory that does not satisfy Hamilton's equations, so we need to resort to the more general residual equation \eqref{SchroedingerPosGeneric}. 

	\bigskip

\begin{figure}
\includegraphics[width=0.99\textwidth]{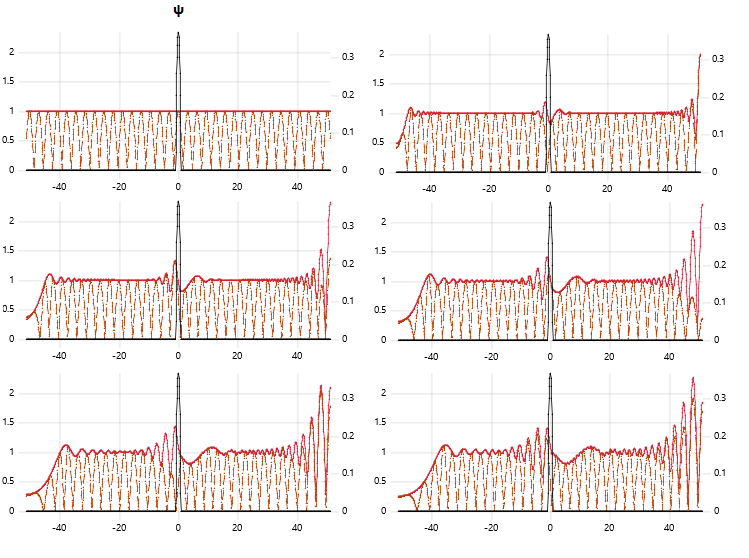}
\caption{Numerical solution of a scattering problem for the Schrödinger equation and a fixed-momentum incoming wave. Timesteps increasing from top left to bottom right.				Neumann boundary conditions have been applied, leading to strong boundary effects that disturb the solution.}
\label{fig:scattering-plot-schroedinger}
\end{figure}

	Consider a 1d potential $V(q)$ localized in some finite interval $I = [q_0, q_1]$. We are interested in the scattering behaviour of incoming waves. Outside the interval $I$, the Schrödinger equation is that of a free particle:
	\begin{equation}
		i\hbar \partial_t \psi = \frac {\hat p^2}{2m}\psi.
	\end{equation}
	Solutions can be written as superpositions of the plane waves

	\begin{equation}\label{planeWave}
		\psi^p(t, x) = \exp\Big\{\frac i\hbar \Big(px - \frac{p^2}{2m}t\Big)\Big\}
	\end{equation}

	for $p \in \mathbb R$. In order to determine the influence of the potential on such a wave, we can solve the Schrödinger equation for the initial condition $\psi(t=0, x) = \psi^p(0,x)$. One important aspect to consider in this case are the boundary conditions, since $\psi^p$ does not vanish rapidly outside some finite domain, as was the case for the other examples considered above. Figure \ref{fig:scattering-plot-schroedinger} illustrates the problems that arise from inappropriate boundary conditions. It shows the results of a numerical integration of the Schrödinger equation with Neumann boundary conditions applied, i.e. $\partial_x \psi| _{\partial \Omega} = 0$, where $\partial \Omega$ denotes the boundary of the considered domain/grid. Reflection artifacts are clearly visible at the left and right boundaries. The potential chosen is 

	\begin{equation}
		V(q) = \exp\Big( -\frac 1{1-q^2} \Big)
	\end{equation}
	
	for $|q| < 1$ and $V(q) = 0$ otherwise. We set $m=1$, $\hbar=1$, $p=1$, and select the time step $\Delta t = 4\pi/1000$. Frames shown in Figure \ref{fig:scattering-plot-schroedinger} are recorded after 0, 150, 300, 450, 600, and 750 integration steps, respectively.

	\bigskip

	Now consider the constant phase space trajectory $c(t) = (0, p)$, with constant momentum $p$ and $q=0$, noting that $c$ does not satisfy Hamilton's equations for a free particle. The resdiual Schrödinger equation \eqref{SchroedingerPosGeneric} for $c$ becomes
	\begin{equation}
		i\hbar \partial_t \Phi(t,x)= \bigg[-\frac {\hbar^2}{2m}\Delta_x  + \frac\hbar i\frac pm \partial_x+ V(x) \bigg] \Phi(t,x).
	\end{equation}
	
\begin{figure}
\includegraphics[width=0.99\textwidth]{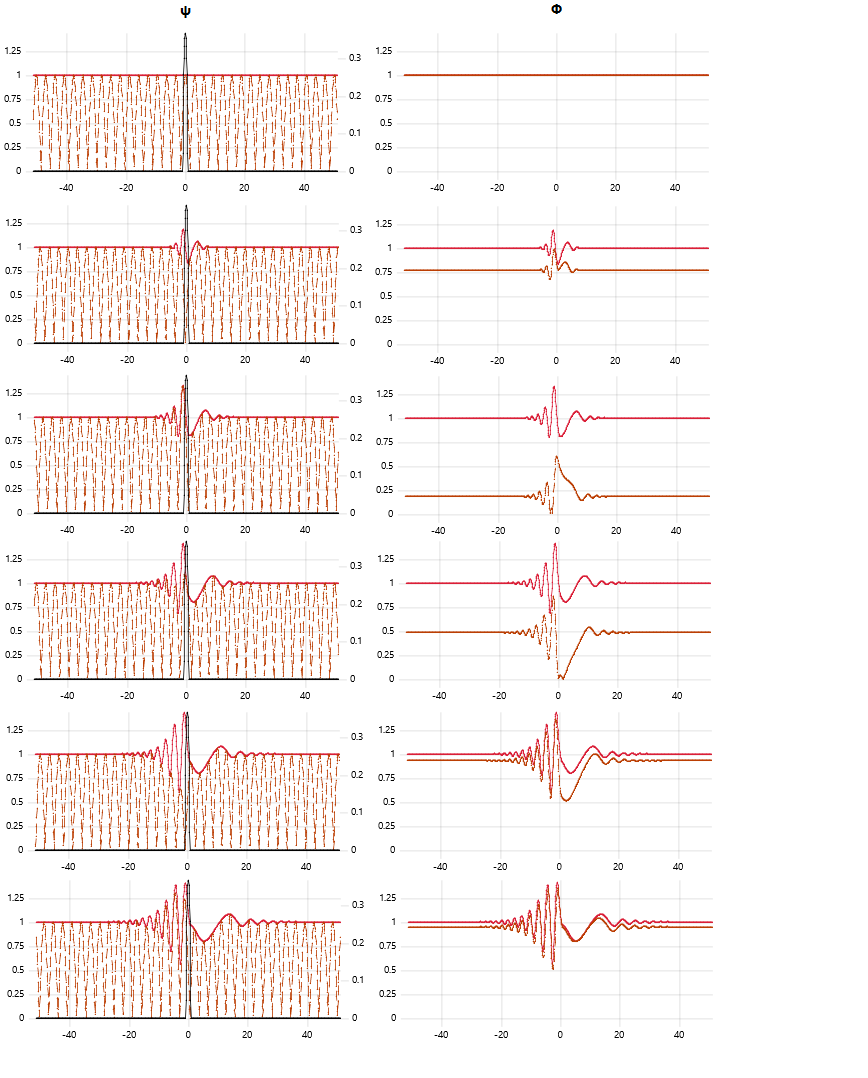}
\caption{Numerical solution of a scattering problem for the Schrödinger equation and a fixed-momentum incoming wave in the residual representation. Timesteps increasing from top to bottom. Neumann boundary conditions in the residual representation are perfectly adapted to the incoming wave, so boundary effects are absent initially.}
\label{fig:scattering-plot-residual}
\end{figure}

	Let us take a look a the residual free particle wave function for momentum $p$:
	\begin{equation}\label{planeWaveTilde}
		\Phi^p(t,x) = U_{c(t)} \psi^p(t,x) = \exp\Big\{\frac i\hbar \Big(\frac{p^2}{2m}t-px\Big)\Big\}\psi^p(x) = 1.
	\end{equation}
	It is spatially constant, since our reference frame already moves with momentum $p$. Therefore, Neumann boundary conditions are perfectly suited for $\Phi^p$. Figure \ref{fig:scattering-plot-residual} shows the result of integrating the problem in the residual representation over $c$, again with Neumann boundary conditions, but this time applied to $\Phi$ instead of $\psi$. We can see that boundary effects are absent in the frames shown. Of course, once the disturbances caused by the potential reach the boundaries, reflections will once again show up.

\section*{Appendix A: Software implementation}\label{sec:appA}\addcontentsline{toc}{section}{Appendix A: Software implementation}

	A prototypical implementation of the numerical solution to the 1-dimensional residual Schrödinger equation has been published alongside this paper at \href{https://github.com/cnoelle/schroedinger-numerics}{https://github.com/cnoelle/schroedinger-numerics} (also available at Zenodo at \href{https://zenodo.org/records/10642346}{https://zenodo.org/records/10642346}). Documentation can be found in the \href{https://github.com/cnoelle/schroedinger-numerics/Readme.md}{\textit{Readme.md}} file in the git repository and guidance on how to reproduce the results shown in the figures of the present work can be found in \href{https://github.com/cnoelle/schroedinger-numerics/Examples.md}{\textit{Examples.md}}.
	 A Hamiltonian of the form $H(q,p) = \frac {p^2}{2m} + V(q)$ is assumed, and a symplectic Euler formulation of Hamilton's equations is used \cite{GaucklerHairerLubich17}: 
	\begin{align}
		p_{n+1} &= p_n - \Delta t_{\text{Ham}} \cdot V'(q_n) \\
		q_{n+1} &= q_n + \tfrac{p_{n+1}}m \Delta t_{\text{Ham}},  \nonumber
	\end{align}
	for the series of phase-space points $(q_n, p_n)_{n=0,1,2,\dots}$ and time step $ \Delta t_{\text{Ham}}$. For the residual Schrödinger equation in position representation \eqref{SchroedingerPosHamiltonClosed} a Crank-Nicolson integration scheme (\cite{CrankNicolson47}) is employed, by solving the matrix equation

	\begin{equation}
		\Big[1+ \frac{ i\Delta t_{\text{S}}}{2\hbar}\big(T + V_{n+1}\big) \Big] \Phi_{n+1} = \Big[1-\frac{i\Delta t_{\text{S}}}{2\hbar}\big(T + V_n\big) \Big] \Phi_n
	\end{equation}

	for the series of wave function $(\Phi_n)_{n=0,1,2, \dots}$. Here $T = \hat p^2 / 2m $ is the kinetic energy operator, $\Delta t_{\text S}$ is the integration time step, and the potential operator is defined by
 
	\begin{equation}
		V_n \cdot \Phi(x) = \big[V(q_n + x) - V(q_n) - x V'(q_n)\big] \Phi(x).
	\end{equation}

	The Schrödinger time step is constrained to an integer multiple of the classical time step, $\Delta t_{\text{S}} = z \Delta t_{\text{Ham}}$ for $z\in \mathbb N$, in the implementation.

\end{document}